\documentclass[epj]{svjour}
\usepackage{graphics}
\usepackage[utf8]{inputenc}
\usepackage{graphicx}
\usepackage{psfrag}
\usepackage{amssymb}
\usepackage{amsmath}
\usepackage{epstopdf}
\usepackage{color}
\usepackage{bm}
\begin{document}
\title{Description of the newly observed $\Xi_c^{0}$ states as molecular states}
\author{HongQiang Zhu\inst{1}
\and NaNa Ma\inst{2}
\and Yin Huang\inst{3}\thanks{\emph{Email address:} huangy2019@swjtu.edu.cn}
%
}                     
%
%
\institute{College of Physics and Electronic Engineering, Chongqing Normal Universit
\and School of Nuclear Science and Technology, Lanzhou University, 730000 Lanzhou, People’s Republic of China
\and School of Physical Science and Technology, Southwest Jiaotong University, Chengdu 610031,China}
\date{Received: date / Revised version: date}
%
\abstract{
Very recently, three new structures $\Xi_c(2923)^0$, $\Xi(2938)^0$, and $\Xi(2964)^0$ at the
invariant mass spectrum of $\Lambda_c^{+}K^{-}$ observed by the LHCb Collaboration triggers a
hot discussion about their inner structure.  In this work, we study the strong decay mode of
the newly observed $\Xi_c$ assuming  that the $\Xi_c$ is a $\bar{D}\Lambda-\bar{D}\Sigma$ molecular
state.  With the possible quantum numbers $J^p=1/2^{\pm}$ and $3/2^{\pm}$, the partial decay
widths of the $\bar{D}\Lambda-\bar{D}\Sigma$ molecular state into the $\Lambda_c^{+}K^{-}$,
$\Sigma_c^{+}K^{-}$,and $\Xi_c^{+}\pi^{-}$ final states through hadronic loop are calculated with
the help of the effective Lagrangians.  By comparison with the LHCb observation, the current results
of total decay width support the $\Xi(2923)^0$ as $\bar{D}\Lambda-\bar{D}\Sigma$ molecule while the
decay width of the $\Xi_c(2938)^0$ and $\Xi(2964)^0$ can not be well reproduced in the molecular state
picture.  In addition, the calculated partial decay widths with $S$ wave $\bar{D}\Lambda-\bar{D}\Sigma$ molecular
state picture indicate that allowed decay modes, $\Xi_c^{+}\pi^{-}$, may have the biggest branching ratios for
the $\Xi_c(2923)$.  The experimental measurements for this strong decay process
could be a crucial test for the molecule interpretation of the $\Xi_c(2923)$. 
} 
\maketitle

\section{Introduction}\label{sec:intro}
During the past several decades,  many narrow baryons with a heavy charm quark, a light up or down quark, and a strange
quark have been reported by the LHCb,CDF Collaboration and so on~\cite{Tanabashi:2018oca}.  Very recently, three other
neutral resonances $\Xi_c^{*0}$ named $\Xi_c(2923)^0$, $\Xi_c(2939)^0$, and $\Xi_c(2965)^0$ have been observed in the
$K^{-}\Lambda_c^{+}$ mass spectra by the LHCb Collaboration~\cite{Aaij:2020yyt}.  The observed resonance
masses and widths are
\begin{align}
&{\rm M}=2923.04\pm{0.25}(stat)\pm0.20(syst)\pm{}0.14(\Lambda^+_c)~ {\rm MeV}\nonumber\\
&\Gamma =7.1\pm0.8(stat)\pm1.8(syst)~{\rm MeV},\nonumber
\end{align}
\begin{align}
&{\rm M}=2938.55\pm{0.21}(stat)\pm0.17(syst)\pm{}0.14(\Lambda^+_c)~ {\rm MeV}\nonumber\\
&\Gamma =10.2\pm0.8(stat)\pm1.1(syst)~{\rm MeV},\nonumber
\end{align}
\begin{align}
&{\rm M}=2964.88\pm{0.26}(stat)\pm0.14\pm{}(syst)0.14(\Lambda^+_c)~ {\rm MeV}\nonumber\\
&\Gamma=14.1\pm0.9(stat)\pm1.3(syst)~{\rm MeV},\nonumber
\end{align}
respectively.  From the observed decay mode, the isospin of these three states are $1/2$.  Although the quantum numbers
of these states remain undetermined, it is very helpful to understand the spectroscopy of the
heavy baryons containing $c$ and $s$ quark.

Due to their observed decay mode, the new structures $\Xi_c(2923)^0$, $\Xi(2938)^0$, and $\Xi(2964)^0$ contain at
least three different valence quark components.  In other word, these states may be candidates of conventional
three-quark state.  Indeed, the QCD sum rule suggests that the newly observed states $\Xi_c(2923)^0$, $\Xi(2938)^0$,
and $\Xi(2964)^0$ are most likely to be considered as the $P$-wave $\Xi_c^{'}$ baryons with the spin-parity $J^p=1/2^{-}$
or $J^p=3/2^{-}$~\cite{Yang:2020zjl}.  In Ref.~\cite{Wang:2020gkn} the $\Xi_c(2923)^0$, $\Xi(2938)^0$, and $\Xi(2964)^0$
ware suggested to be $1P$ $\Xi_c^{'}$ state with spin-parity $J^P = 3/2^-$ or $J^P = 5/2^-$  in the chiral quark model.
In Ref.~\cite{Lu:2020ivo} the two-body strong decays of the $\Xi_c(2923)^0$, $\Xi(2938)^0$, and $\Xi(2964)^0$ were calculated by
employing the $^3P_0$ approach with the conclusion that the $\Xi_c(2923)^0$ and $\Xi(2938)^0$ can be $1P$ $\Xi_c^{'}$ states,
and the $\Xi(2964)^0$ can be regarded as the $2S$ $\Xi_c^{'}$ state.  The lattice QCD calculation was also performed and
try to determine their quantum numbers~\cite{Bahtiyar:2020uuj}.

Although the authors in Refs.~\cite{Yang:2020zjl,Wang:2020gkn,Lu:2020ivo,Bahtiyar:2020uuj} try to assign these states into the
conventional three-quark frames,  it is obvious that the inner structure of $\Xi_c(2923)^0$, $\Xi(2938)^0$, and $\Xi(2964)^0$ are
not finally determined.   And another interpretation is treating them as $D\Lambda-D\Sigma$ molecular states, because the smallest mass
gaps between the newly observed $\Xi_c$ baryons and the ground $\Xi_c$, about 450 MeV, is large enough to excite a light quark-antiquark
pair to form a hadronic molecular.  Indeed, it is shown in Refs.~\cite{JimenezTejero:2009vq,Yu:2018yxl,Nieves:2019jhp} that the
interaction between $D$ meson and $\Lambda$ or $\Sigma$ baryon is strong enough to form a bound state with a mass about 2930 MeV.

The key point in this work is to explain whether the $\Xi_c(2923)^0$, $\Xi(2938)^0$, and $\Xi(2964)^0$ can be considered as a molecular state.
Here, we will consider the strong $\Xi_c^{*}\to\Lambda_c^+K^{-}$,$\Sigma_c^+K^{-}$, and $\Xi_c^{+}\pi^{-}$ decays of the
$\Xi_c(2923)^0$, \\$\Xi(2938)^0$, and $\Xi(2964)^0$ with the possible quantum numbers $J^p=1/2^{\pm}$ and $3/2^{\pm}$ using an effective
Lagrangian approach.   The approach is based on the hypothesis that the $\Xi_c^{*}$ is a hadronic molecule state of $D\Lambda$-$D\Sigma$.
The coupling of the $\Xi_c^{*}$ to the constituents is described by the effective Lagrangian.  The corresponding coupling constant $g_{\Xi_c^{*}\Lambda{}D}$ and $g_{\Xi_c^{*}\Sigma{}D}$ are determined by the compositeness condition $Z=0$~\cite{Faessler:2007gv,Faessler:2007us,Dong:2009tg,Dong:2010xv,Dong:2010gu}, which implies that the renormalization constant of the hadron wave function is set equal to zero.   By constructing a phenomenological Lagrangian including
the couplings of the bound state to its constituents and the constituents with other particles we calculated one-loop diagrams describing different decays
of the molecular states.

This work is organized as follows. The theoretical
formalism is explained in Sec.~\ref{section:2}. The predicted partial
decay widths are presented in Sec.~\ref{section:3}, followed by a short
summary in the last section.

\section{FORMALISM AND INGREDIENTS}\label{section:2}
\begin{figure*}[htbp]
\begin{center}
\includegraphics[scale=0.50]{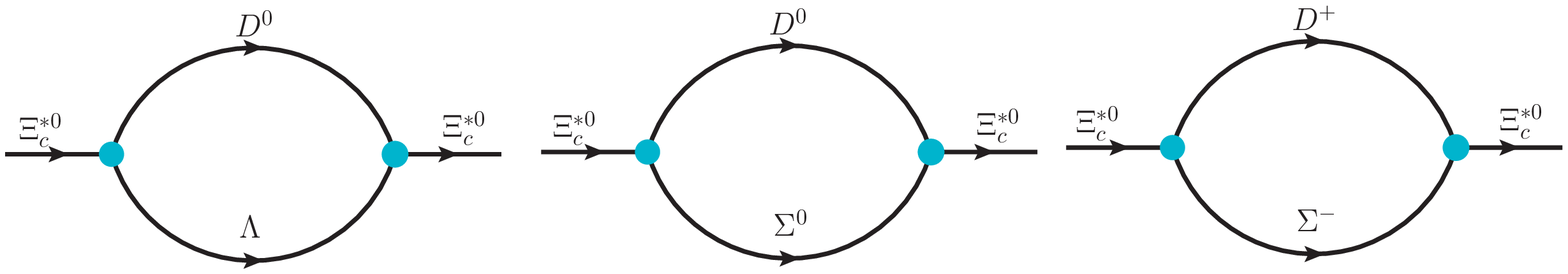}
\caption{Self-energy of the $\Xi_c^{*}$ states..}\label{mku2}
\end{center}
\end{figure*}
In the molecular scenario, the details of the calculations for $\Xi_c^{0}\to\Lambda_c^{+}K^{-}$, $\Xi_c^{0}\to\Sigma_c^{+}K^{-}$, and $\Xi_c^{0}\to\Xi_c^{+}\pi^{-}$ are
presented for $\Xi_c$ state with two different total angular momentum $J$ possibilities.    The molecular structure of the $\Xi_c^{*}$
baryon with quantum numbers $J^p=1/2^{\pm}$ is described by the Lagrangian~\cite{Dong:2010gu}
\begin{align}
{\cal{L}}_{\Xi_c^{*}}(x)&=g_{\Xi_c^{*}D\Lambda}\int{}d^4y\Phi(y^2)D^0(x+\omega_{\Lambda}y)\Gamma\Lambda(x-\omega_{D^0}y)\nonumber\\
                        &\times\bar{\Xi}_c^{*}(x)+g_{\Xi_c^{*}D\Sigma}\int{}d^4y\Phi(y^2)[\sqrt{\frac{2}{3}}D^0(x+\omega_{\Sigma^0}y)\nonumber\\
                        &\times\Gamma\Sigma^0(x-\omega_{D^0}y)-\sqrt{\frac{1}{3}}D^{+}(x+\omega_{\Sigma^{-}}y)\nonumber\\
                        &\times\Gamma\Sigma^{-}(x-\omega_{D^{+}}y)]\bar{\Xi}_c^{*}(x)\label{eq1}.
\end{align}
while for the choice $J^p=3/2^{\pm}$ the Lagrangian contains a derivative
\begin{align}
{\cal{L}}_{\Xi_c^{*}}(x)&=g_{\Xi_c^{*}D\Lambda}\int{}d^4y\Phi(y^2)D^0(x+\omega_{\Lambda}y)\Gamma\partial_{\mu}\Lambda(x-\omega_{D^0}y)\nonumber\\
                        &\times\bar{\Xi}_c^{*\mu}(x)+g_{\Xi_c^{*}D\Sigma}\int{}d^4y\Phi(y^2)[\sqrt{\frac{2}{3}}D^0(x+\omega_{\Sigma^0}y)\nonumber\\
                        &\times\Gamma\partial_{\mu}\Sigma^0(x-\omega_{D^0}y)-\sqrt{\frac{1}{3}}D^{+}(x+\omega_{\Sigma^{-}}y)\nonumber\\
                        &\times\Gamma\partial_{\mu}\Sigma^{-}(x-\omega_{D^{+}}y)]\bar{\Xi}_c^{*\mu}(x)\label{eq2}.
\end{align}
where $\Gamma$ is the corresponding Dirac matrix related to the spin-parity of the $\Xi_c^{*}$.  In particular, for
$J^{p}=1/2^{+},3/2^{-}$ we have $\Gamma=\gamma^{5}$ while for $J^{p}=1/2^{-},3/2^{+}$ the Dirac structure $\Gamma=1$.
In the Lagrangian, an effective correlation function $\Phi(y^2)$ is introduced to reflect the distribution of two
constituents in the hadronic molecular $\Xi_c^{*}$ state.   The introduced correlation function also makes the
Feynman diagrams finite in the ultraviolet region of Euclidean space, which indicates that the Fourier transformation
of the correlation function should drop fast enough in the ultraviolet region. Here we choose the Fourier transformation of the
correlation in the Gaussian form~\cite{Faessler:2007gv,Faessler:2007us,Dong:2009tg,Dong:2010xv,Dong:2010gu},
\begin{align}
\Phi(p_E^2)\doteq\exp(-p_E^2/\alpha^2)
\end{align}
with $\alpha$ being the size parameter which characterize the distribution of components inside the molecule. Though the value of $\alpha$
could not be determined from first principles, it is usually chosen to be about $1$ GeV
in the literature~\cite{Faessler:2007gv,Faessler:2007us,Dong:2009tg,Dong:2010xv,Dong:2010gu}.  In this work, we set $\Lambda=1.0$ GeV.

With the help of the effective Lagrangian in Eq.(~\ref{eq1}) and Eq.(~\ref{eq2}), we can obtain the self energy of the $\Xi_c^{*}$
\begin{align}
\Sigma^{1/2}_{\Xi_c^{*}}(k_0)&=\int\frac{d^4k_1}{(2\pi)^4}\{g^2_{\Xi_c^{*}\Lambda{D}}\Phi^2[(k_1-k_0\omega_{\Lambda})^2]\Gamma\frac{k\!\!\!/_1+m_{\Lambda}}{k_1^2-m^2_{\Lambda}}\Gamma\nonumber\\
                             &\times\frac{1}{(k_1-k_0)^2-m^2_{D^{0}}}+g^2_{\Xi_c^{*}\Sigma{}D}[\frac{2}{3}\Phi^2[(k_1-k_0\omega_{\Sigma^0})^2]\nonumber\\
                             &\times\Gamma\frac{k\!\!\!/_1+m_{\Sigma^0}}{k_1^2-m^2_{\Sigma^0}}\Gamma\frac{1}{(k_1-k_0)^2-m^2_{D^{0}}}+\frac{1}{3}\Phi^2[(k_1-k_0\omega_{\Sigma^{-}})^2]\nonumber\\
                             &\times\Gamma\frac{k\!\!\!/_1+m_{\Sigma^{-}}}{k_1^2-m^2_{\Sigma^{-}}}\Gamma\frac{1}{(k_1-k_0)^2-m^2_{D^{+}}}]\}\\
\Sigma^{3/2}_{\Xi_c^{*}}(k_0)&=\int\frac{d^4k_1}{(2\pi)^4}\{g^2_{\Xi_c^{*}\Lambda{D}}\Phi^2[(k_1-k_0\omega_{\Lambda})^2]\Gamma\frac{k\!\!\!/_1+m_{\Lambda}}{k_1^2-m^2_{\Lambda}}\Gamma\nonumber\\
                             &\times{}\frac{1}{(k_1-k_0)^2-m^2_{D^{0}}}+g^2_{\Xi_c^{*}\Sigma{}D}[\frac{2}{3}\Phi^2[(k_1-k_0\omega_{\Sigma^0})^2]\nonumber\\
                             &\times\Gamma\frac{k\!\!\!/_1+m_{\Sigma^0}}{k_1^2-m^2_{\Sigma^0}}\Gamma\frac{1}{(k_1-k_0)^2-m^2_{D^{0}}}+\frac{1}{3}\Phi^2[(k_1-k_0\omega_{\Sigma^{-}})^2]\nonumber\\
                             &\times\Gamma\frac{k\!\!\!/_1+m_{\Sigma^{-}}}{k_1^2-m^2_{\Sigma^{-}}}\Gamma\frac{1}{(k_1-k_0)^2-m^2_{D^{+}}}]\}k_1^{\mu}k_1^{\nu}
\end{align}
where $k_0^2=m^2_{\Xi_c^{*}}$ with $k_0, m^{*}_{\Xi_c}$ denoting the four momenta and mass of the $\Xi_c^{*}$, respectively,
$k_1$, $m_{D}$, and $m_{\Lambda}$, $m_{\Sigma}$ are the four-momenta, the mass of the $D$ meson, and the mass of the $\Lambda$
baryon, the mass of the $\Sigma$ baryon, respectively.   The coupling constant $g_{\Xi_c^{*}\Lambda\bar{D}}$ and $g_{\Xi_c^{*}\Sigma{}D}$
is determined by the compositeness condition~\cite{Weinberg:1962hj,Salam:1962ap}. It implies that the renormalization constant
of the hadron wave function is set equal to zero with
\begin{align}
Z_{\Xi_c^{*}}=x_{D\Lambda}+x_{D\Sigma}-\frac{d\Sigma^{1/2,3/2-T}_{\Xi_c^{*}}}{dk\!\!\!/_0}|_{k\!\!\!/_0=m_{\Xi^{*}_{c}}}=0.
\end{align}
where the $x_{AB}$ is the probability to find the $\Xi_c^*$ in the hadronic state $AB$ with the normalization $x_{D\Lambda}+x_{D\Sigma}=1.0$.
And the $\Sigma^{3/2-T}_{\Xi_c^{*}}$ is the transverse part of the self-energy operator $\Sigma^{3/2}_{\Xi_c^{*}}$, related to
$\Sigma^{3/2}_{\Xi_c^{*}}$ via
\begin{align}
\Sigma^{3/2}_{\Xi_b^{*}}(k_0)=(g_{\mu\nu}-\frac{k_0^{\mu}k_0^{\nu}}{k_0^2})\Sigma^{3/2-T}_{\Xi_c^{*}}+\cdots.\label{eq3}
\end{align}

Fig.~(\ref{fety}) shows the hadronic decay of the $\Lambda{}D-\Sigma{D}$ molecular state into the $\Lambda_c^{+}K^{-}$,$\Xi_c^{+}\pi^{-}$, and $\Sigma_c^{+}K^{-}$ final states occuring by exchanging nucleon, $D_s^{*}$ meson, and $D^{*}$ meson.   To compute the amplitudes of the diagrams
shown in Fig.(~\ref{fety}),  we need the effective Lagrangian densities for the relevant interaction vertices.
In Refs.~\cite{Hofmann:2005sw,Montana:2017kjw}, coupling of the vector meson to charm baryons are described from effective Lagrangians, which are obtained
using the hidden gauge formalism and assuming SU(4) symmetry:
\begin{figure}[htbp]
\begin{center}
\includegraphics[bb=40 150 650 710, clip,scale=0.45]{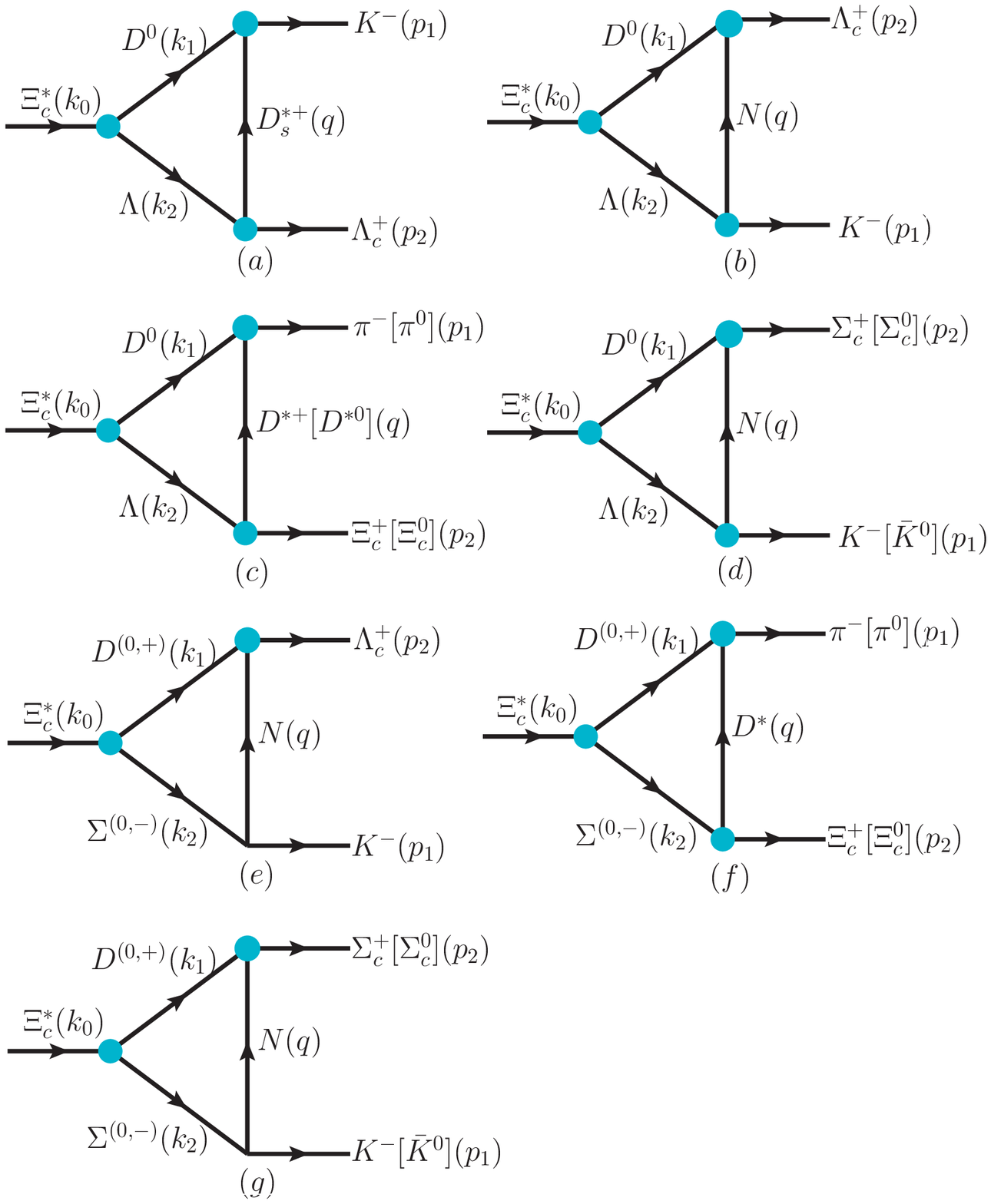}
\end{center}
\caption{Feynman diagrams for the $\Xi_c^{*}\to{}\Lambda_c^{+}K^{-}$,$\Xi_c\pi$, and $\Sigma_c\bar{K}$ decay process. }\label{fety}
\end{figure}
\begin{align}
{\cal{L}}_{VBB}=\frac{g_1}{4}\sum_{i,j,k,l=1}^{4}\bar{B}_{ijk}\gamma^{\mu}(V^{K}_{\mu,l}B^{ijl}+2V^{j}_{\mu,l}B^{ilk})\label{eq8},
\end{align}
where the coupling constant $g=6.6$ is  from Ref.~\cite{Hofmann:2005sw}.  The symbol $V_{\mu}$ represents the vector fields of the 16-plet of the $\rho$, given by
\begin{equation}
V_{\mu}=
\left(
  \begin{array}{cccc}
    \frac{1}{\sqrt{2}}(\rho^{0}+\omega) & \rho^{+}                             &  K^{*+}     & \bar{D}^{*0} \\
    \rho^{-}                            & \frac{1}{\sqrt{2}}(-\rho^{0}+\omega) &  K^{*0}     & D^{*-}       \\
     K^{*-}                             & \bar{K}^{*0}                         &  \phi       & D^{*-}_s     \\
     D^{*0}                             & D^{*+}                               &  D^{*+}_{s} & J/\psi       \\
  \end{array}
\right)_{\mu},
\end{equation}
and $B$ is the tensor of baryons belonging to the 20-plet of $p$
\begin{align}
&B^{121}=p,                        ~~~~~B^{122}=n,                                        ~~~~~ B^{132}=\frac{1}{\sqrt{2}}\Sigma^{0}-\frac{1}{\sqrt{6}}\Lambda,\nonumber\\
&B^{213}=\sqrt{\frac{2}{3}}\Lambda,~~~~~B^{231}=\frac{1}{\sqrt{2}}\Sigma^{0}+\frac{1}{\sqrt{6}}\Lambda,~~~~~B^{232}=\Sigma^{-},\nonumber\\
&B^{233}=\Xi^{-},B^{311}=\Sigma^{+},~~~~~B^{313}=\Xi^{0},~~~~~B^{141}=-\Sigma_c^{++},\nonumber\\
&B^{142}=\frac{1}{\sqrt{2}}\Sigma^{+}_{c}+\frac{1}{\sqrt{6}}\Lambda_c,~~~~~B^{143}=\frac{1}{\sqrt{2}}\Xi^{'+}_c-\frac{1}{\sqrt{6}}\Xi_c^{+},\nonumber\\
&B^{241}=\frac{1}{\sqrt{2}}\Sigma_c^{+}-\frac{1}{\sqrt{6}}\Lambda_c,~~~~~B^{242}=\Sigma_c^{0},\nonumber\\
&B^{243}=\frac{1}{\sqrt{2}}\Xi^{'0}_c+\frac{1}{\sqrt{6}}\Xi_c^{0},~~~~~B^{341}=\frac{1}{\sqrt{2}}\Xi^{,+}_c+\frac{1}{\sqrt{6}}\Xi_c^{+},\nonumber\\
&B^{124}=\sqrt{\frac{2}{3}}\Lambda_c,~~~~~B^{234}=\sqrt{\frac{2}{3}}\Xi_c^{0},~~~~~B^{314}=\sqrt{\frac{2}{3}}\Xi_c^{+},\nonumber\\
&B^{342}=\frac{1}{\sqrt{2}}\Xi^{'0}_c-\frac{1}{\sqrt{6}}\Xi_c^{0},~~~~~B^{343}=\Omega_c^{0},\nonumber\\
&B^{144}=\Xi^{++}_{cc},~~~~~B^{244}=-\Xi^{+}_{cc},~~~~~B^{344}=\Omega_{cc}\label{eqw2},
\end{align}
where the indices $i,j,k$ of $B^{ijk}$ denote the quark content of the baryon fields with the identification $1\leftrightarrow{u},2\leftrightarrow{d},3\leftrightarrow{}s$,and $4\leftrightarrow{}c$.

To evaluate the diagrams in Fig.(~\ref{fety}), in addition to the Lagrangian in Eq.(~\ref{eq1}), Eq.(~\ref{eq2}), and Eq.(~\ref{eq8}),
the following effective Lagrangians, responsible for vector mesons and pseudoscalar mesons interactions are needed
as well~\cite{Hofmann:2005sw}
\begin{align}
{\cal{L}}_{PPV}=\frac{i}{4}g_{h}\langle[\partial^{\mu}P,P]V_{\mu}\rangle\label{eq9},
\end{align}
where $P$ is the $SU(4)$ pseudoscalar meson matrices, and $\langle...\rangle$ in the trace over the $SU(4)$ matrices.  The meson matrices are~\cite{Hofmann:2005sw}
\begin{equation}
P=\sqrt{2}
\left(
  \begin{array}{cccc}
    \frac{\pi^{0}}{\sqrt{2}}+\frac{\eta}{\sqrt{6}}+\frac{\eta^{'}}{\sqrt{3}}   &\pi^{+}    &K^{+}      & D^0    \\
    \pi^{-}  &  -\frac{\pi^{0}}{\sqrt{2}}+\frac{\eta}{\sqrt{6}}+\frac{\eta^{'}}{\sqrt{3}}  &K^{0}      &-D^{-}  \\
     K^{-}   &    \bar{K}^{0}  & -\sqrt{\frac{2}{3}}\eta+\frac{\eta^{'}}{\sqrt{3}}                     & D^{-}_s\\
     D^0     &    -D^{+}       &D_s^{+}                                                                &\eta_c   \\
  \end{array}
\right)\label{eqw9}.
\end{equation}
The coupling $g_h$ is fixed from the strong decay width of $K^{*}\to{}K\pi$.  With the help of Eq.~(\ref{eqw9}), the two-body decay width
 $\Gamma(K^{*+}\to{}K^{0}\pi^{+})$ is related to $g_h$ as
 \begin{align}
 \Gamma(K^{*+}\to{}K^{0}\pi^{+})=\frac{g_h^2}{6\pi{}m^2_{K^{*+}}}{\cal{P}}^3_{\pi{}K^{*}}=\frac{2}{3}\Gamma_{K^{*+}},
 \end{align}
where ${\cal{P}}_{\pi{}K^{*}}$ is the three-momentum of the $\pi$ in the rest frame of the  $K^{*}$.
Using the experimental strong decay width($\Gamma_{K^{*+}}=50.3\pm0.8$ MeV) and the masses of the particles needed in
the present work are listed in Table.~(\ref{table1}) we obtain $g=4.64$~\cite{Tanabashi:2018oca}.
\begin{table}[h!]
\centering
\caption{Masses of the particles needed in the present work (in units of MeV).}\label{table1}
\begin{tabular}{cccccccccccccc}
\hline\hline
   $\Lambda$    &$\Lambda_c^{+}$  &$\Sigma_c^{+}$  &$\Sigma_c^{0}$  &$\Xi_c^{+}$    &$D^0$         \\
   $1115.683$   &$2286.46$        &$2452.90$       &$2453.75$       &$2467.93$      &$1864.83$     \\ \hline
   $p$          &$n$              &$K^{0}$         &$K^{\pm}$       &$\pi^{\pm}$    &$D^{\pm}$     \\
   $938.272$    &$939.565$        &$497.611$       &$493.68$        &$139.57$       &$1869.65$     \\  \hline
   $\Xi_c^{0}$  &$D_s^{*\pm}$     &$D^{*\pm}$      &$D^{*0}$        &$\Sigma^{+}$   &$\Sigma^{0}$  \\
   $2470.91$    &$2112.1$         &$2010.26$       &$2006.85$       &$1189.37$      &$1192.642$     \\  \hline  \hline
\end{tabular}
\end{table}

The vertexes for the meson-baryon interaction are needed and the form in the $SU(3)$ sector is given by the chiral Lagrangian~\cite{Garzon:2012np}
\begin{align}
{\cal{L}}_{\phi{\cal{B}}{\cal{B}}}&=\frac{F}{2}\langle\bar{\cal{B}}\gamma_{\mu}\gamma_5[u^{\mu},{\cal{B}}]\rangle+\frac{D}{2}\langle\bar{\cal{B}}\gamma_{\mu}\gamma_5\{u^{\mu},{\cal{B}}\}\rangle\label{Pbb},
\end{align}
where $F=0.51$,$D=0.75$~\cite{Garzon:2012np} and at lowest order in the pseudoscalar field $u^{\mu}=-\sqrt{2}\partial^{\mu}\phi/f$, with $f=93$ MeV.
And ${\cal{B}}$ and $\phi$ is now the $SU(3)$ matrix of the baryon octet and meson, respectively.
\begin{equation}
{\cal{B}}=
\left(
  \begin{array}{ccc}
    \frac{1}{\sqrt{2}}\Sigma^{0}+\frac{1}{\sqrt{6}}\Lambda  & \Sigma^{+}                                               &  p      \\
     \Sigma^{-}                                             & -\frac{1}{\sqrt{2}}\Sigma^{0}+\frac{1}{\sqrt{6}}\Lambda  &  n       \\
     \Xi^{-}                                                & \Xi^{0}                                                  &  -\frac{2}{\sqrt{6}}\Lambda     \\
  \end{array}
\right).
\end{equation}
\begin{equation}
\phi=
\left(
  \begin{array}{ccc}
    \frac{\pi^{0}}{\sqrt{2}}+\frac{\eta}{\sqrt{6}} &    \pi^{+}                                        &       K^{+}\\
    \pi^{-}                                       &    -\frac{\pi^{0}}{\sqrt{2}}+\frac{\eta}{\sqrt{6}} &       K^{0}\\
    K^{-}                                         &    \bar{K}^{0}                                     &       -\frac{2}{\sqrt{6}}\eta
  \end{array}
\right)\label{eq7}.
\end{equation}

Moreover, the effective Lagrangians for the $DN\Lambda_c$ and $DN\Sigma_c$ couplings are~\cite{Xie:2015zga,Huang:2016ygf}
\begin{align}
&{\cal{L}}_{DN\Lambda_c}=ig_{\Lambda_cpD}\bar{\Lambda}_c\gamma_5pD^0,\\
&{\cal{L}}_{DN\Sigma_c}=-ig_{\Sigma_cpD}\bar{N}\gamma_5\vec{\tau}\cdot\vec{\Sigma}_cD,
\end{align}
with $\vec{\tau}$ being the usual Pauli matrices.   The coupling constant $g_{\Lambda_cpD}=10.7^{+5.3}_{-4.3}$ and $g_{\Sigma_cpD}=-2.69$ are
borrowed from Refs.~\cite{Xie:2015zga,Huang:2016ygf,Khodjamirian:2011sp}, where we take the central
values $g_{\Lambda_cpD}=10.7$ in our calculation.

With the above vertices, the amplitudes of the triangle diagrams of Fig.(~\ref{fety}), evaluated in the center of mass frame of
final states, are
\begin{align}
{\cal{M}}_{a}&=(i)^3\frac{g_1g_hg_{\Xi_c^{*}\Lambda{D^0}}}{2}\int\frac{d^4q}{(2\pi)^4}\Phi[(k_1\omega_{\Lambda}-k_2\omega_{D})^2]\nonumber\\
             &\times\bar{u}(p_2)\gamma_{\mu}\frac{(k\!\!\!/_2+m_{\Lambda})}{k_2^2-m^2_{\Lambda}}\Gamma{}\{u(k_0),ik_{2\rho}u^{\rho}(k_0)\}\nonumber\\
             &\times{}\frac{1}{k^2_1-m^2_{D^0}}(p_1^{\nu}+k_1^{\nu})\frac{(-g^{\mu\nu}+q^{\mu}q^{\nu}/m^2_{D_s^{*+}})}{q^2-m^2_{D_s^{*+}}},\nonumber
             \end{align}
\begin{align}
{\cal{M}}_{b}&=-(i)^3\frac{(D+3F)g_{\Lambda_cpD}g_{\Xi_c^{*}\Lambda{D^0}}}{2\sqrt{3}f}\int\frac{d^4q}{(2\pi)^4}\nonumber\\
             &\times\Phi[(k_1\omega_{\Lambda}-k_2\omega_{D})^2]\bar{u}(p_2)\frac{(q\!\!\!/-m_{N})}{q^2-m^2_{N}}p\!\!\!/_1\nonumber\\
             &\times{}\frac{k\!\!\!/_2+m_{\Lambda}}{k_2^2-m^2_{\Lambda}}\Gamma{}\{u(k_0),ik_{2\rho}u^{\rho}(k_0)\}\frac{1}{k^2_1-m^2_{D^0}},\nonumber\\
{\cal{M}}_{c}&=-(i)^3\frac{g_1g_hg_{\Xi_c^{*}\Lambda{D^0}}}{4}\int\frac{d^4q}{(2\pi)^4}\Phi[(k_1\omega_{\Lambda}-k_2\omega_{D})^2]\nonumber\\
              &\times{}\bar{u}(p_2)\gamma_{\mu}\frac{(k\!\!\!/_2+m_{\Lambda})}{k_2^2-m^2_{\Lambda}}\Gamma\{u(k_0),ik_{2\rho}u^{\rho}(k_0)\}\nonumber\\
             &\times{}\frac{1}{k^2_1-m^2_{D^0}}(p_1^{\nu}+k_1^{\nu})\frac{(-g^{\mu\nu}+q^{\mu}q^{\nu}/m^2_{D^{*+}})}{q^2-m^2_{D^{*+}}},\nonumber\\
{\cal{M}}_{d}&=-(i)^3\frac{(D+3F)g_{\Sigma_cpD}g_{\Xi_c^{*}\Lambda{D^0}}}{2\sqrt{3}f}\int\frac{d^4q}{(2\pi)^4}\nonumber\\
             &\times\Phi[(k_1\omega_{\Lambda}-k_2\omega_{D})^2]\bar{u}(p_2)\frac{(q\!\!\!/-m_{N})}{q^2-m^2_{N}}p\!\!\!/_1\nonumber\\
             &\times{}\frac{k\!\!\!/_2+m_{\Lambda}}{k_2^2-m^2_{\Lambda}}\Gamma\{u(k_0),ik_{2\rho}u^{\rho}(k_0)\}\frac{1}{k^2_1-m^2_{D^0}},\nonumber\\
{\cal{M}}_{e1}&=(i)^3\frac{(F-D)g_{\Lambda_cpD}g_{\Xi_c^{*}\Sigma^0D^0}}{2\sqrt{3}f}\int\frac{d^4q}{(2\pi)^4}\nonumber\\
             &\times\Phi[(k_1\omega_{\Sigma^0}-k_2\omega_{D^0})^2]\bar{u}(p_2)\frac{(q\!\!\!/-m_{p})}{q^2-m^2_{p}}p\!\!\!/_1\nonumber\\
             &\times{}\frac{k\!\!\!/_2+m_{\Sigma^0}}{k_2^2-m^2_{\Sigma^0}}\Gamma\{u(k_0),ik_{2\rho}u^{\rho}(k_0)\}\frac{1}{k^2_1-m^2_{D^0}},\nonumber\\
{\cal{M}}_{e2}&=-(i)^3\frac{(F-D)g_{\Lambda_cpD}g_{\Xi_c^{*}\Sigma^-D^+}}{\sqrt{6}f}\int\frac{d^4q}{(2\pi)^4}\nonumber\\
              &\times\Phi[(k_1\omega_{\Sigma^-}-k_2\omega_{D^+})^2]\bar{u}(p_2)\frac{(q\!\!\!/-m_{n})}{q^2-m^2_{n}}p\!\!\!/_1\nonumber\\
             &\times{}\frac{k\!\!\!/_2+m_{\Sigma^-}}{k_2^2-m^2_{\Sigma^-}}\Gamma\{u(k_0),ik_{2\rho}u^{\rho}(k_0)\}\frac{1}{k^2_1-m^2_{D^+}},\nonumber\\
{\cal{M}}_{f1}&=(i)^3\frac{g_1gg_{\Xi_c^{*}\Sigma^0{D^0}}}{4}\int\frac{d^4q}{(2\pi)^4}\Phi[(k_1\omega_{\Sigma^0}-k_2\omega_{D^0})^2]\nonumber\\
             &\times{}\bar{u}(p_2)\gamma_{\mu}\frac{(k\!\!\!/_2+m_{\Sigma^0})}{k_2^2-m^2_{\Sigma^0}}\Gamma\{u(k_0),ik_{2\rho}u^{\rho}(k_0)\}\frac{1}{k^2_1-m^2_{D^0}}\nonumber\\
             &\times(p_1^{\nu}+k_1^{\nu})\frac{(-g^{\mu\nu}+q^{\mu}q^{\nu}/m^2_{D^{*+}})}{q^2-m^2_{D^{*+}}},\nonumber\\
{\cal{M}}_{f2}&=-(i)^3\frac{g_1gg_{\Xi_c^{*}\Sigma^-{D^+}}}{2}\int\frac{d^4q}{(2\pi)^4}\Phi[(k_1\omega_{\Sigma^-}-k_2\omega_{D^+})^2]\nonumber\\
              &\times{}\bar{u}(p_2)\gamma_{\mu}\frac{(k\!\!\!/_2+m_{\Sigma^-})}{k_2^2-m^2_{\Sigma^-}}\Gamma\{u(k_0),ik_{2\rho}u^{\rho}(k_0)\}\frac{1}{k^2_1-m^2_{D^+}}\nonumber\\
             &\times{}(p_1^{\nu}+k_1^{\nu})\frac{(-g^{\mu\nu}+q^{\mu}q^{\nu}/m^2_{D^{*0}})}{q^2-m^2_{D^{*0}}},\nonumber\\
{\cal{M}}_{g1}&=-(i)^3\frac{(F-D)g_{\Sigma_cpD}g_{\Xi_c^{*}\Sigma^0{D^0}}}{\sqrt{6}f}\int\frac{d^4q}{(2\pi)^4}\nonumber\\
              &\times\Phi[(k_1\omega_{\Sigma^0}-k_2\omega_{D})^2]\bar{u}(p_2)\frac{(q\!\!\!/-m_{N})}{q^2-m^2_{N}}p\!\!\!/_1\nonumber\\
             &\times{}\frac{k\!\!\!/_2+m_{\Sigma^0}}{k_2^2-m^2_{\Sigma^0}}\Gamma\{u(k_0),ik_{2\rho}u^{\rho}(k_0)\}\frac{1}{k^2_1-m^2_{D^0}},\nonumber
\end{align}
\begin{align}{\cal{M}}_{g2}&=(i)^3\frac{\sqrt{6}(F-D)g_{\Sigma_cpD}g_{\Xi_c^{*}\Sigma^-{D^+}}}{3f}\int\frac{d^4q}{(2\pi)^4}\nonumber\\
             &\times{}\Phi[(k_1\omega_{\Sigma^-}-k_2\omega_{D^+})^2]\bar{u}(p_2)\frac{(q\!\!\!/-m_{N})}{q^2-m^2_{N}}\nonumber\\
             &\times{}p\!\!\!/_1\frac{k\!\!\!/_2+m_{\Sigma^-}}{k_2^2-m^2_{\Sigma^-}}\Gamma\{u(k_0),ik_{2\rho}u^{\rho}(k_0)\}\nonumber\\
             &\times\frac{1}{k^2_1-m^2_{D^+}}.
\end{align}

Once the amplitudes are determined, the corresponding partial decay widths can be obtained, which read,
\begin{align}
\Gamma(\Xi_c^{*}\to MB)=\frac{1}{2J+1}\frac{1}{8\pi}\frac{|\vec{p}_1|}{m^2_{\Xi_c^{*}}}\overline{|{\cal{M}}|^2},
\end{align}
where $J$ is the total angular momentum of the $\Xi_b^{*}$ state, the $|\vec{p}_1|$ is the three-momenta of the decay
products in the center of mass frame, the overline indicates the sum over the polarization vectors of the final hadrons,
and $MB$ denotes the decay channel of $MB$, i.e.,$\Lambda_c\bar{K}$, $\Xi_c\pi$, $\Sigma_c\bar{K}$.

\section{RESULTS AND DISCUSSIONS}\label{section:3}
\begin{figure}[htbp]
\begin{center}
\includegraphics[bb=40 75 650 600, clip,scale=0.45]{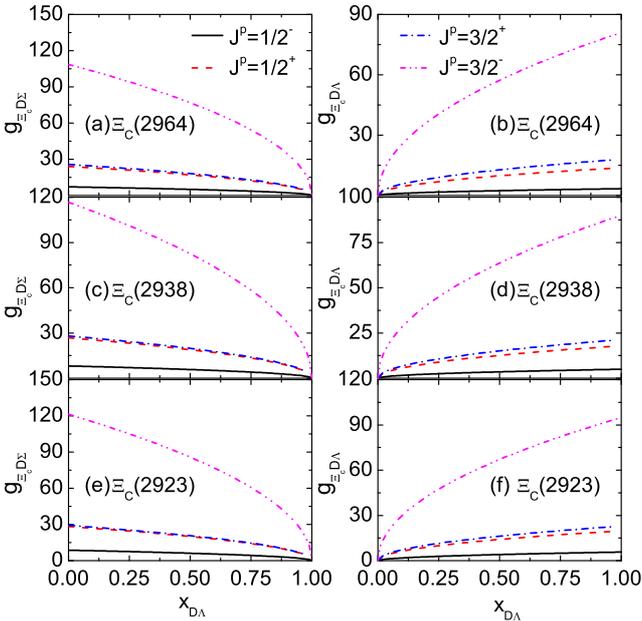}
\end{center}
\caption{The coupling constants of the $\Xi_c^*$ state with different $J^P$
assignments as a function of the parameter $x_{D\Lambda}$.  And the $x_{D\Lambda}$
is the probability to find the $\Xi_c^{*}$ in the hadronic
state $D\Lambda$. }\label{feydiagrams}
\end{figure}
In this work, the main ideal is to explain whether the $\Xi_c(2923)^0$, $\Xi(2938)^0$, and $\Xi(2964)^0$ can be considered
as a $D\Lambda-D\Sigma$ molecular state by computing allowed two-body strong decay modes.  To compute the partial decay widths of the
$\Xi_c^{*}$, we first need the information of the coupling constants related to the molecular state and its components.  In
Fig.~\ref{feydiagrams} we show the results for the coupling constant $g_{\Xi_c^*\Lambda{}D}$ and $g_{\Xi_c^*\Sigma{}D}$ of the
various $\Xi_c^*$ states for different spin-parity assignments and for a variation of the size parameter $x_{D\Lambda}$ in a range
of $0.0-1.0$.   In the discussed $x_{D\Lambda}$ range, the coupling constant $g_{\Xi_c^*\Sigma{}D}$ decreases, while the coupling
constant $g_{\Xi_c^*\Lambda{}D}$ increases.   The results in Fig.~\ref{feydiagrams} also show that  the coupling constants
increases (or decreases) sharply  with the increase of the $x_{D\Lambda}$ for the $J^p=3/2^{-}$ case, where the $\Xi_c^*$ is a
$D$-wave $D\Lambda-D\Sigma$ molecular state.  For the  case of $J^p=1/2^{-}$, the coupling constants increases (or decreases)
but relatively slowly compared with that coupling constants in the case of $J^p=3/2^{-}$.

\begin{figure}[htbp]
\begin{center}
\includegraphics[bb=55 40 680 510, clip,scale=0.38]{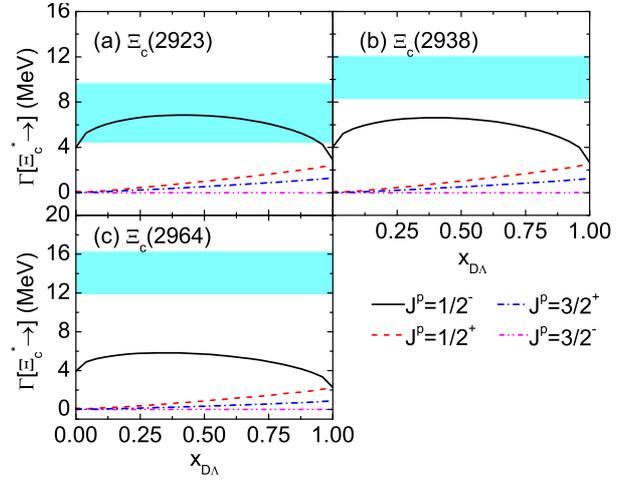}
\end{center}
\caption{The total decay width  with different spin-parity assignments for the various $\Xi_c^{*}$ as a function of the
parameter $x_{D\Lambda}$.  The cyan bands denote the experimental total width.~\cite{Aaij:2020yyt}.
 }\label{decay-width}
\end{figure}
We show the dependence of the total decay width on the $x_{D\Lambda}$ in Fig.~\ref{decay-width}.   The total decay
widths increase with $x_{D\Lambda}$ for the $J^P=1/2^{+}$ and $J^P=3/2^{\pm}$ assignments.  For the $J^P=1/2^-$ assignments,
we found that the line shape of the the total decay widths are huge different and the total decay widths first increases,
then decreases but very slowly.   A possible explanation for this may be that for an $S-$wave loosely bound state the
effective coupling strength of the bound state to its components is more sensitive to the inner structure than the
effective coupling strength of another possible molecular state, such as $P-$ wave molecular state, relative to their
inner component.  This is why people often focus on the bound state from $S$-wave interaction and assume the $P-$ and $D$-wave
bound state should be difficult to form from hadron-hadron interaction~\cite{He:2016pfa}.

From the Fig.~\ref{decay-width}, one find that the predicted total decay widths for the $\Xi_c(2964)$ state and $\Xi_c(2938)$ state
in the four spin-parity assignments are all smaller than the experimental total width.   Such results disfavor the
assignment of these two states as $D\Lambda-D\Sigma$ molecular state.  For the $\Xi_c(2923)$ state, the predicted total decay width is
much smaller than the experimental total width in the case of $J^P=1/2^{+}$, which disfavors such a spin-parity assignment for the $\Xi_c(2923)$
in the $D\Lambda-D\Sigma$ molecular picture.  For the case of $J^P=3/2^{+}$,  Since the estimated total decay widths is much smaller than
the experimental total width, this case can be completely excluded as well.   The $J^P=3/2^-$ case is  also disfavored
due to the smallest width predicted.   Hence, only the $\Xi_{c}(2923)$ can be considered as the molecular states composed of $D\Lambda$ and $D\Sigma$ components by comparison with the total decay width experimentally measured.   The results in Fig.~\ref{decay-width} also show that the total
decay width of the $\Xi_c(2923)$ can not be reproduced when only consider $\Xi_c(2923)$ as pure $D\Lambda$ or pure $D\Sigma$ molecular state.

\begin{figure}[htbp]
\begin{center}
\includegraphics[bb=35 30 680 520, clip,scale=0.30]{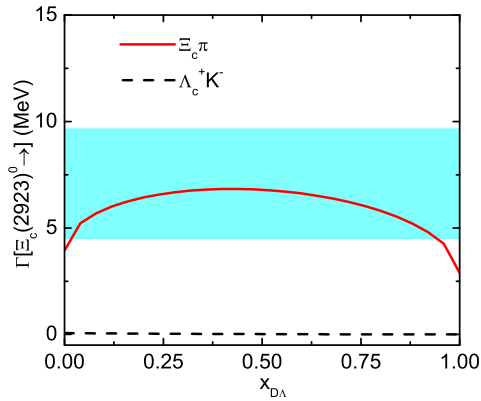}
\end{center}
\caption{Partial decay widths of $\Xi_c(2923)^0\to\Lambda_c^+K^-$ (red solid line) and $\Xi_c(2923)^0\to\Xi_c\pi$ (black dash line)
 with $J^P=1/2^{-}$ as a function of the parameter $x_{D\Lambda}$.  The cyan bands denote the experimental total width.~\cite{Aaij:2020yyt}.
 }\label{indiv-decay-width}
\end{figure}
The partial decay widths of $\Xi_c(2923)^0\to\Lambda_c^+K^-$ and $\Xi_c(2923)^0\to\Xi_c\pi$ with $J^P=1/2^{-}$ assignment as a function of the
parameter $x_{D\Lambda}$ are presented in Fig.~\ref{indiv-decay-width}.  It is found that the transition $\Xi_c(2923)^0\to\Xi_c\pi$ is main
decay channel, which almost saturates the total width of $\Xi_c(2923)$.  However, the transition $\Xi_c(2923)^0\to\Lambda_c^+K^-$  give minor
contributions.   The decay width $\Xi_c(2923)^0\to\Xi_c\pi$ is very different from that in the constituent quark model~\cite{Yang:2020zjl,Wang:2020gkn,Lu:2020ivo} if we assign the $S$-wave $D\Lambda-D\Sigma$ bound state as $\Xi_c(2923)$.
Future experimental
measurements of such a process can be quite useful to test
the different interpretations of the $\Xi_c(2923)$.

\section{Summary}\label{sec:summary}

In this work, the $S$-wave $D\Lambda-D\Sigma$ molecular states were studied by calculating their allowed two body strong decay to investigate whether
the three new narrow $\Xi_c^*$ baryons, $\Xi^{*}(2923)$, $\Xi_c^*(2938)$, and $\Xi_c^*(2964)$ can be understood as $D\Lambda-D\Sigma$ molecules.
With the coupling constants obtained by the composition condition, the decays through hadronic loops are calculated in a phenomenological
effective Lagrangian approach. The total decay widths can be well reproduced with the assumption that the $\Xi^{*}(2923)$ as $S$-wave $D\Lambda-D\Sigma$
bound state with $J^P=1/2^{-}$, which decay channels are  $\Lambda_c^+K^-$ and $\Xi_c\pi$.   The other newly reported $\Xi_c^*$ states
cannot be accommodated in the current molecular picture.  If the $\Xi^{*}(2923)$ is pure $D\Lambda-D\Sigma$  molecule, the transition strength of $\Xi_c(2923)^0\to\Xi_c\pi$ is quite different from that in the constituent quark model~\cite{Yang:2020zjl,Wang:2020gkn,Lu:2020ivo} and the decay
width  almost saturates the total width of $\Xi_c(2923)$.  Future experimental
measurements of such a process can be quite useful to test
the different interpretations of the $\Xi_c(2923)$.

\section*{Acknowledgments}
This work is partly supported by the Development and Exchange Platform for Theoretic Physics of
Southwest Jiaotong University in 2020(Grants No.11947404).  We acknowledge the supported by the National
Science Foundation of Chongqing (Grant No. cstc2019jcyj-msxm0953), the Science and Technology Research
Program of Chongqing Municipal Education Commission (Grant No. KJQN201800510).  NaNa Ma also acknowledge
the supported by the National Natural Science Foundation of China(No. 11947229) and Postdoctoral Science
Foundation of China (No. 2019M663853)

\end{document}